\documentclass[12pt,a4paper]{article}

\usepackage{subfigure}
\usepackage{lineno}  % for line numbering during review
\usepackage{graphicx}  % to include figures (can also use other packages)
\usepackage{xspace} % To avoid problems with missing or double spaces after
\usepackage{color}
\usepackage{colortbl}
\usepackage{amsmath} % Adds a large collection of math symbols
\usepackage{ifthen} % for conditional statements

\newboolean{pdflatex}
\setboolean{pdflatex}{true} % use this if using non-eps figures
\newboolean{articletitles}
\setboolean{articletitles}{true} % False removes titles in references

\newboolean{uprightparticles}
\setboolean{uprightparticles}{false} %Set to true to get roman particle symbols

\usepackage{amssymb}
\usepackage{amsfonts}
\usepackage{upgreek} % Adds in support for greek letters in roman typeset

\usepackage{hyperref}    % Hyperlinks in references
\usepackage[all]{hypcap} % Internal hyperlinks to floats.

\def\lhcb {LHCb\xspace}
\def\ux85 {UX85\xspace}

\def\babar  {BaBar\xspace}
\def\belle  {Belle\xspace}

\ifthenelse{\boolean{uprightparticles}}%
{

 \def\PDelta      {\ensuremath{\Delta}\xspace}                 
 \def\PXi      {\ensuremath{\Xi}\xspace}                 
 \def\PLambda      {\ensuremath{\Lambda}\xspace}                 
 \def\PSigma      {\ensuremath{\Sigma}\xspace}                 
 \def\POmega      {\ensuremath{\Omega}\xspace}                 
 \def\PUpsilon      {\ensuremath{\Upsilon}\xspace}                 
 
 \def\PB      {\ensuremath{\mathrm{B}}\xspace}                 
                  
 \def\PD      {\ensuremath{\mathrm{D}}\xspace}

 \def\PK      {\ensuremath{\mathrm{K}}\xspace}

 \def\Pb      {\ensuremath{\mathrm{b}}\xspace}                 
 \def\Pc      {\ensuremath{\mathrm{c}}\xspace}

 \def\Pi      {\ensuremath{\mathrm{i}}\xspace}

 \def\Ps      {\ensuremath{\mathrm{s}}\xspace}

}
{

 \mathchardef\PDelta="7101
 \mathchardef\PXi="7104
 \mathchardef\PLambda="7103
 \mathchardef\PSigma="7106
 \mathchardef\POmega="710A
 \mathchardef\PUpsilon="7107
                  
 \def\PB      {\ensuremath{B}\xspace}                 
                  
 \def\PD      {\ensuremath{D}\xspace}

 \def\PK      {\ensuremath{K}\xspace}

 \def\Pb      {\ensuremath{b}\xspace}                 
 \def\Pc      {\ensuremath{c}\xspace}

 \def\Pi      {\ensuremath{i}\xspace}

 \def\Ps      {\ensuremath{s}\xspace}

}

\def\squark    {\ensuremath{\Ps}\xspace}

\def\cquark    {\ensuremath{\Pc}\xspace}

\def\bquark    {\ensuremath{\Pb}\xspace}

\def\kaon  {\ensuremath{\PK}\xspace}
  \def\Kbar  {\kern 0.2em\overline{\kern -0.2em \PK}{}\xspace}

\def\Kz    {\ensuremath{\kaon^0}\xspace}
\def\Kzb   {\ensuremath{\Kbar^0}\xspace}
\def\KzKzb {\ensuremath{\Kz \kern -0.16em \Kzb}\xspace}
\def\Kp    {\ensuremath{\kaon^+}\xspace}
\def\Km    {\ensuremath{\kaon^-}\xspace}

\def\KpKm  {\ensuremath{\Kp \kern -0.16em \Km}\xspace}

\def\Kstarz  {\ensuremath{\kaon^{*0}}\xspace}
\def\Kstarzb {\ensuremath{\Kbar^{*0}}\xspace}

  \def\Dbar    {\kern 0.2em\overline{\kern -0.2em \PD}{}\xspace}
\def\D       {\ensuremath{\PD}\xspace}

\def\Dz      {\ensuremath{\D^0}\xspace}
\def\Dzb     {\ensuremath{\Dbar^0}\xspace}
\def\DzDzb   {\ensuremath{\Dz {\kern -0.16em \Dzb}}\xspace}
\def\Dp      {\ensuremath{\D^+}\xspace}
\def\Dm      {\ensuremath{\D^-}\xspace}

\def\DpDm    {\ensuremath{\Dp {\kern -0.16em \Dm}}\xspace}

\def\B       {\ensuremath{\PB}\xspace}
  \def\Bbar    {\kern 0.18em\overline{\kern -0.18em \PB}{}\xspace}

\def\Bd      {\ensuremath{\B^0}\xspace}
\def\Bs      {\ensuremath{\B^0_\squark}\xspace}
\def\Bsb     {\ensuremath{\Bbar^0_\squark}\xspace}

  \def\Y#1S{\ensuremath{\PUpsilon{(#1S)}}\xspace}% no space before {...}!

\def\to                 {\ensuremath{\rightarrow}\xspace}

\def\CP                {\ensuremath{C\!P}\xspace}
\def\CPT               {\ensuremath{C\!PT}\xspace}

\newcommand{\DGs}{\ensuremath{\Delta\Gamma_{\squark}}\xspace}

\newcommand{\Gs}{\ensuremath{\Gamma_{\squark}}\xspace}

\newcommand{\GL}{\ensuremath{\Gamma_{\rm L}}\xspace}
\newcommand{\GH}{\ensuremath{\Gamma_{\rm H}}\xspace}

\newcommand{\phis}{\ensuremath{\phi_{\squark}}\xspace}

\def\AT#1     {\ensuremath{A_{\mathrm{T}}^{#1}}\xspace}           % 2

\def\C#1      {\ensuremath{\mathcal{C}_{#1}}\xspace}                       % 9
\def\Cp#1     {\ensuremath{\mathcal{C}_{#1}^{'}}\xspace}                    % 7
\def\Ceff#1   {\ensuremath{\mathcal{C}_{#1}^{\mathrm{(eff)}}}\xspace}        % 9  
\def\Cpeff#1  {\ensuremath{\mathcal{C}_{#1}^{'\mathrm{(eff)}}}\xspace}       % 7
\def\Ope#1    {\ensuremath{\mathcal{O}_{#1}}\xspace}                       % 2
\def\Opep#1   {\ensuremath{\mathcal{O}_{#1}^{'}}\xspace}                    % 7

\newcommand{\unit}[1]{\ensuremath{\rm\,#1}\xspace}          % {kg}

\newcommand{\tev}{\ensuremath{\mathrm{\,Te\kern -0.1em V}}\xspace}
\newcommand{\gev}{\ensuremath{\mathrm{\,Ge\kern -0.1em V}}\xspace}
\newcommand{\mev}{\ensuremath{\mathrm{\,Me\kern -0.1em V}}\xspace}
\newcommand{\kev}{\ensuremath{\mathrm{\,ke\kern -0.1em V}}\xspace}
\newcommand{\ev}{\ensuremath{\mathrm{\,e\kern -0.1em V}}\xspace}
\newcommand{\gevc}{\ensuremath{{\mathrm{\,Ge\kern -0.1em V\!/}c}}\xspace}
\newcommand{\mevc}{\ensuremath{{\mathrm{\,Me\kern -0.1em V\!/}c}}\xspace}
\newcommand{\gevcc}{\ensuremath{{\mathrm{\,Ge\kern -0.1em V\!/}c^2}}\xspace}
\newcommand{\gevgevcccc}{\ensuremath{{\mathrm{\,Ge\kern -0.1em V^2\!/}c^4}}\xspace}
\newcommand{\mevcc}{\ensuremath{{\mathrm{\,Me\kern -0.1em V\!/}c^2}}\xspace}

\def\mum  {\ensuremath{\,\upmu\rm m}\xspace}

\def\invfb   {\ensuremath{\mbox{\,fb}^{-1}}\xspace}

\def\ps   {\ensuremath{{\rm \,ps}}\xspace}

\newcommand{\chisq}{\ensuremath{\chi^2}\xspace}

\def\gsim{{~\raise.15em\hbox{$>$}\kern-.85em
          \lower.35em\hbox{$\sim$}~}\xspace}
\def\lsim{{~\raise.15em\hbox{$<$}\kern-.85em
          \lower.35em\hbox{$\sim$}~}\xspace}

\def\pt         {\mbox{$p_{\rm T}$}\xspace}

\def\gauss      {\mbox{\textsc{Gauss}}\xspace}

\def\tell1  {TELL1\xspace}
\def\ukl1   {UKL1\xspace}

\newcommand{\phisphiphi}{\phis}
\newcommand{\particle}[1]{{\ensuremath{\rm #1}}}

\newcommand{\DLL}[2]{\ensuremath{\Delta  \ln{\cal L}_{\particle{#1}\particle{#2}}}}
\newcommand{\MeVc}{\unit{MeV\!/\!{\it c}}}
\newcommand{\MeVcc}{\unit{MeV\!/\!{\it c}^2}}

\usepackage{cite}
\usepackage{mciteplus}

\begin{document}

%%%%%%%%%%%%%%%%%%%%%%%%%
%%%%% Title     %%%%%%%%%
%%%%%%%%%%%%%%%%%%%%%%%%%

%%%%%%%%%%%%%%%%%%%%%%%%%
%%%%%  TITLE PAGE  %%%%%%
%%%%%%%%%%%%%%%%%%%%%%%%%
\begin{titlepage}
\pagenumbering{roman}

% Header ---------------------------------------------------
\vspace*{-1.5cm}
\centerline{\large EUROPEAN ORGANIZATION FOR NUCLEAR RESEARCH (CERN)}
\vspace*{1.5cm}
\hspace*{-0.5cm}
\begin{tabular*}{\linewidth}{lc@{\extracolsep{\fill}}r}
\ifthenelse{\boolean{pdflatex}}% Logo format choice
{\vspace*{-2.7cm}\mbox{\!\!\!\includegraphics[width=.14\textwidth]{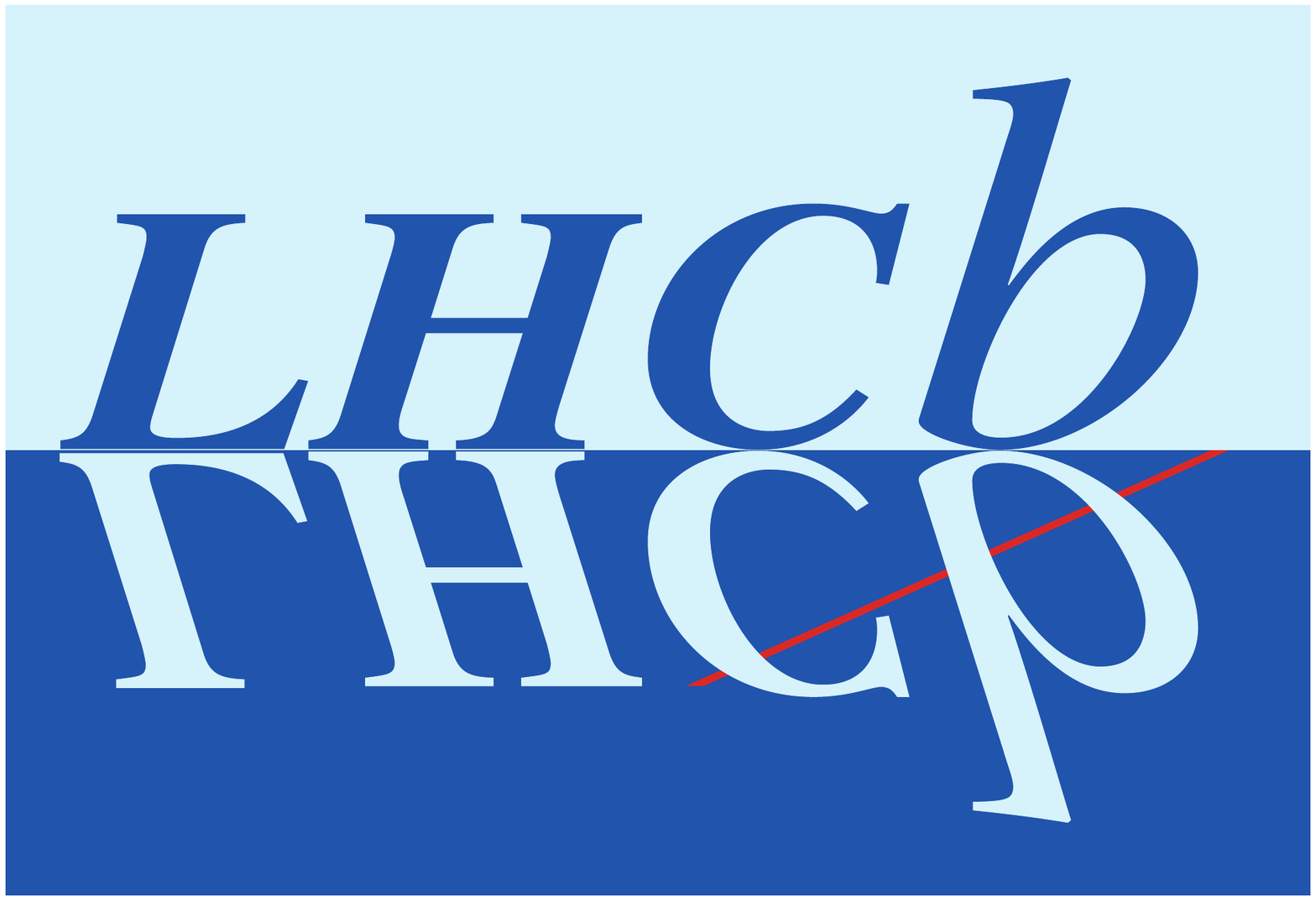}} & &}%
{\vspace*{-1.2cm}\mbox{\!\!\!\includegraphics[width=.12\textwidth]{lhcb-logo.eps}} & &}%
\\
 & & CERN-PH-EP-2012-092 \\  % ID 
 & & LHCb-PAPER-2012-004 \\  % ID 
 & & June 15, 2012 \\ % Date - Can also hardwire e.g.: 23 March 2010
 & & \\
\end{tabular*}

\vspace*{4.0cm}

% Title --------------------------------------------------
{\bf\boldmath\huge
\begin{center}
  Measurement of the polarization amplitudes and triple product
  asymmetries in the $\Bs \rightarrow \phi \phi$ decay
\end{center}
}

\vspace*{1.0cm}

% Authors -------------------------------------------------
\begin{center}
The LHCb collaboration
\footnote{Authors are listed on the following pages.}
\end{center}

\vspace{\fill}

% Abstract -----------------------------------------------
\begin{abstract}
  \noindent
Using $1.0~\invfb$ of $pp$ collision data collected at a centre-of-mass
energy of $\sqrt{\mathrm{s}}=7~{\mathrm{TeV}}$ with the LHCb detector,
measurements of the polarization amplitudes, strong phase difference and triple product
asymmetries in the $\Bs \rightarrow \phi \phi$ decay mode are
presented. The measured values are
\begin{alignat*}{3}
|A_0|^2 &=& 0.365 \pm 0.022 \,({\rm stat}) \pm 0.012  \,({\rm syst}) \, , \\
|A_\perp|^2 &=& 0.291 \pm 0.024 \,({\rm stat}) \pm 0.010  \,({\rm syst}) \, , \\
\cos(\delta_\parallel) &=&  -0.844 \pm  0.068 \,({\rm stat}) \pm 0.029  \,({\rm syst}) \, , \\
A_U &=& -0.055 \pm 0.036 \,({\rm stat}) \pm 0.018  \,({\rm syst}) \, , \\
A_V &=& 0.010 \pm 0.036 \,({\rm stat}) \pm 0.018  \,({\rm syst}) \, . 
\end{alignat*}
\end{abstract}

\end{titlepage}

%%%%%%%%%%%%%%%%%%%%%%%%%%%%%%%%
%%%%%  EOD OF TITLE PAGE  %%%%%%
%%%%%%%%%%%%%%%%%%%%%%%%%%%%%%%%

%  empty page follows the title page ----
\newpage
\setcounter{page}{2}
\mbox{~}
\newpage

% Author List ----------------------------
%  You need to get a new author list!
%%%%%%%%%%%%%%%%%%%%%%%%%%%%%%%%%%%%%%%%%%
\centerline{\large\bf LHCb collaboration}
\begin{flushleft}
\small
R.~Aaij$^{38}$, 
C.~Abellan~Beteta$^{33,n}$, 
B.~Adeva$^{34}$, 
M.~Adinolfi$^{43}$, 
C.~Adrover$^{6}$, 
A.~Affolder$^{49}$, 
Z.~Ajaltouni$^{5}$, 
J.~Albrecht$^{35}$, 
F.~Alessio$^{35}$, 
M.~Alexander$^{48}$
% ...

{\footnotesize \it
% ...
$ ^{5}$Clermont Universit\'{e}, Universit\'{e} Blaise Pascal, CNRS/IN2P3, LPC, Clermont-Ferrand, France\\
$ ^{6}$CPPM, Aix-Marseille Universit\'{e}, CNRS/IN2P3, Marseille, France\\
% ...
$ ^{33}$Universitat de Barcelona, Barcelona, Spain\\
$ ^{34}$Universidad de Santiago de Compostela, Santiago de Compostela, Spain\\
$ ^{35}$European Organization for Nuclear Research (CERN), Geneva, Switzerland\\
% ...
$ ^{38}$Nikhef National Institute for Subatomic Physics, Amsterdam, The Netherlands\\
% ...
$ ^{43}$H.H. Wills Physics Laboratory, University of Bristol, Bristol, United Kingdom\\
% ...
$ ^{48}$School of Physics and Astronomy, University of Glasgow, Glasgow, United Kingdom\\
$ ^{49}$Oliver Lodge Laboratory, University of Liverpool, Liverpool, United Kingdom\\
% ...
$ ^{n}$LIFAELS, La Salle, Universitat Ramon Llull, Barcelona, Spain\\
% ...
}
\end{flushleft}
%%%%%%%%%%%%%%%%%%%%%%%%%%%%%%%%%%%%%%%%%%

\cleardoublepage

%%%%%%%%%%%%%%%%%%%%%%%%%%%%%%%%
%%%%%  Table of Content   %%%%%%
%%%%%%%%%%%%%%%%%%%%%%%%%%%%%%%%
%%%% Uncomment next 2 lines if desired
%\cleardoublepage

%%%%%%%%%%%%%%%%%%%%%%%%%
%%%%% Main text %%%%%%%%%
%%%%%%%%%%%%%%%%%%%%%%%%%

\pagestyle{plain} % restore page numbers for the main text
\setcounter{page}{1}
\pagenumbering{arabic}

% %%%%%%% CHOOSE --------
%% ----------------------------------
%% Line numbering on the left margin 
%% ----------------------------------
%% Uncomment during review phase. 
%% Comment it out before a final submission.
%\linenumbers
%% --------------------------------
% %%%%%%%%%%%%% ---------

%  You can include short sections directly in the main tex file. 
%  However, for larger papers it is desirable to split 
%  the text into several semiautonomous files, which can be revised independently. 
%  This is especially useful when developing a document in collaboration with several people, 
%  since then different parts can be edited independently. 
%  This type of file organization is shown here. 
% 

\section{Introduction}
\label{sec:Introduction}
In the Standard Model, the flavour-changing neutral current
decay $\Bs \rightarrow \phi \phi$ proceeds via a $b\rightarrow
s\bar{s}s$ penguin process. Studies of the polarization amplitudes and triple product asymmetries  
in this decay provide
powerful tests for the presence of contributions from 
processes beyond the Standard Model~\cite{gronau, Bensalem:2000hq, Datta:2003mj, Nandi:2005kh, Datta:2011qz}.

The  $\Bs \rightarrow \phi \phi$ decay is a pseudoscalar to vector-vector transition. As a result, there are three possible spin
configurations of the vector meson pair allowed by angular momentum 
conservation. These manifest themselves as three helicity states, with
amplitudes denoted $H_{+1}, H_{-1}$ and $H_0$. It is convenient to define linear polarization amplitudes,
which are related to the helicity amplitudes through the following
transformations
\begin{eqnarray}
A_{0}  &=&  H_{0} \, , \nonumber \\
A_\perp &=& \frac{H_{+1} - H_{-1}}{\sqrt{2}} \, ,   \nonumber \\
A_\parallel &=& \frac{H_{+1} + H_{-1}}{\sqrt{2}} \, . 
\end{eqnarray}

The $\phi\phi$ final state can be a mixture of \CP-even and \CP-odd eigenstates. 
The longitudinal ($A_0$) and parallel ($A_\parallel$) components are \CP-even and 
the perpendicular component ($A_\perp$) is \CP-odd.
From the V--A structure of the weak interaction, the longitudinal component, $f_L = |A_0|^2/(|A_0|^2+|A_\perp|^2+|A_\parallel|^2)$, is
expected to be dominant \cite{beneke, PhysRevD.80.114026,
  PhysRevD.76.074018}. However, roughly equal longitudinal and
transverse components are found in measurements of $B^{+} \rightarrow \phi K^{*+}$, $B^0 \rightarrow \phi K^{*0}$,
$B^{+} \rightarrow \rho^0 K^{*+}$ and $B^0 \rightarrow \rho^0 K^{*0}$
decays at the B-factories~\cite{PhysRevLett.94.221804,PhysRevLett.98.051801,PhysRevD.78.092008, delAmoSanchez:2010mz,Abe:2004mq, Aubert:2006fs}.
To explain this, large contributions
from either penguin annihilation effects \cite{kagan} or final state
interactions \cite{PhysRevD.76.034015} have been proposed. Recent calculations where
phenomenological parameters are adjusted to account for the data allow $f_L$ in the range $0.4 - 0.7$ \cite{beneke, PhysRevD.80.114026}.  
Another pseudoscalar to vector-vector penguin decay is $\Bs \to \Kstarz \Kstarzb$. A recent measurement by the LHCb Collaboration in this decay mode has found a value of $f_L=0.31\pm0.12\pm0.04$ \cite{kstarlhcb}. 

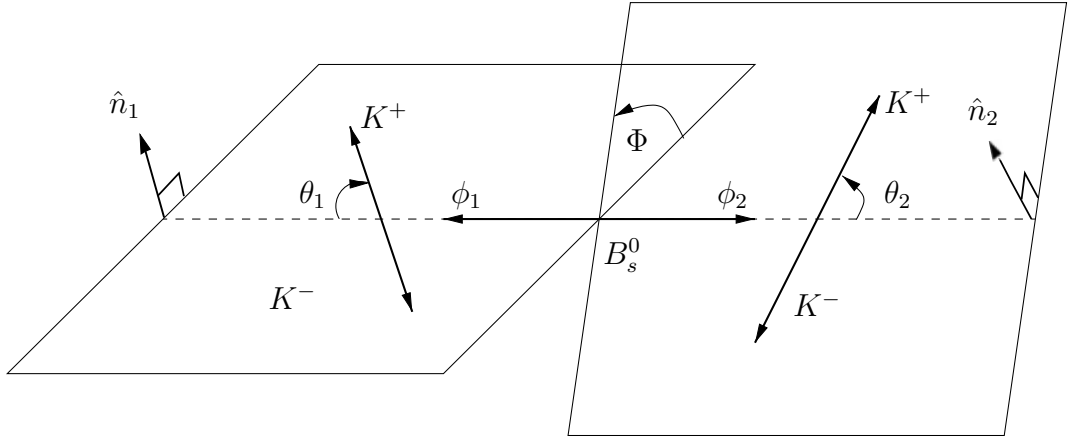
\begin{figure}[ht]
\setlength{\unitlength}{1mm}
  \centering
  \begin{picture}(140,60) %90
    \put(0,-1){
      \includegraphics*[width=140mm,%height=100mm,%
      ]{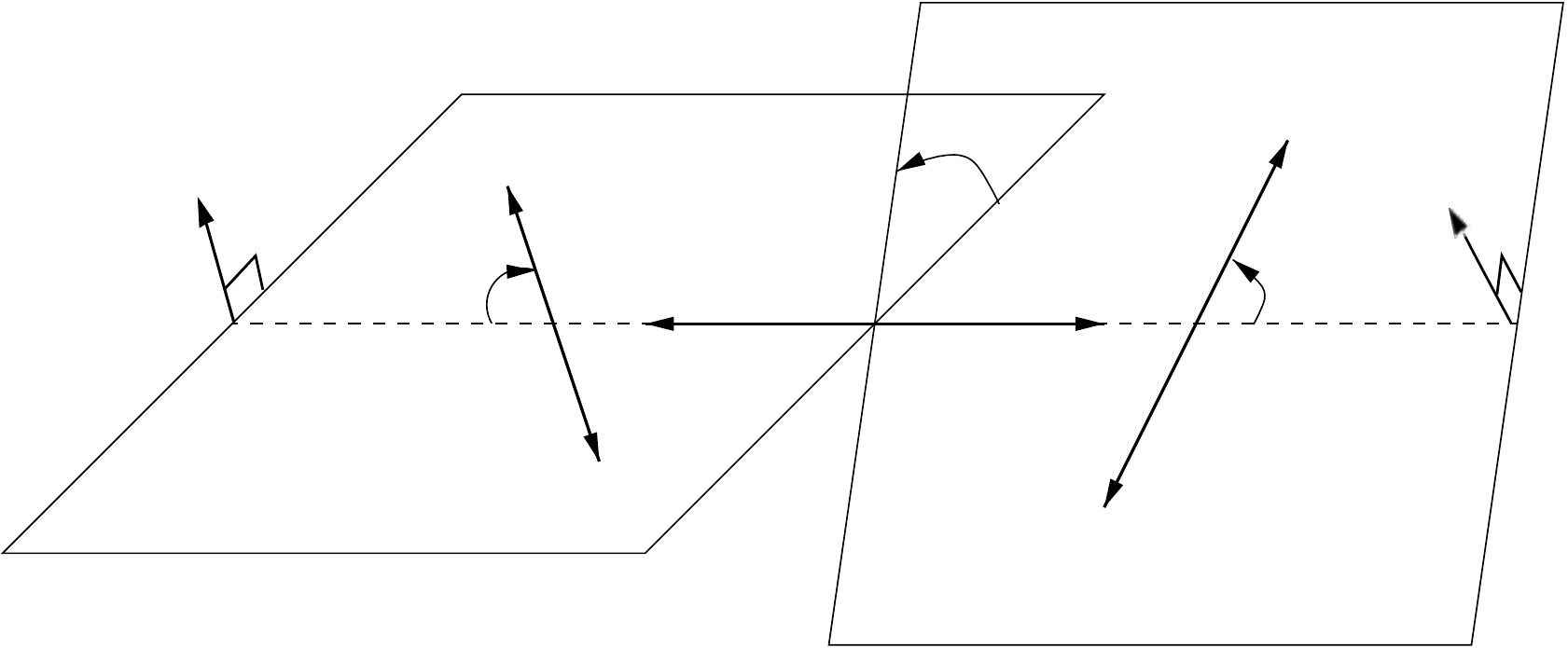}
    }
      \put(15,42){$\hat{n}_{1}$}
      \put(128,41){$\hat{n}_{2}$}
      \put(80,22){$B^0_{s}$}
      \put(83,37){$\Phi$}
      \put(117,30){$\theta_{2}$}
      \put(40,30){$\theta_{1}$}
      \put(36,16){$K^-$}
      \put(48,40){$K^+$}
      \put(105,15){$K^-$}
      \put(117,42){$K^+$}
      \put(60,30){$\phi_1$}
      \put(95,30){$\phi_2$}   
  \end{picture}
  \caption { 
    Decay angles for the $\Bs \rightarrow \phi \phi$ decay, where the $K^+$ momentum in the $\phi_{1,2}$ rest frame,
    and the parent $\phi_{1,2}$ momentum in the rest frame of the \Bs meson span the two $\phi$ meson decay planes,
    $\theta_{1,2}$ is the angle between the $K^+$ track momentum in the $\phi_{1,2}$ meson rest frame and
    the parent $\phi_{1,2}$ momentum in the \Bs rest frame, $\Phi$ is the angle between the two $\phi$ meson
    decay planes and $\hat{n}_{1,2}$ is the unit vector normal to the decay plane of the $\phi_{1,2}$ meson.
  }
  \label{decay_angles} 
\end{figure}

The time-dependent differential decay rate for the $\Bs \rightarrow \phi \phi$ mode can be written as
\begin{equation}
\frac{d^4 \Gamma}{d\cos \theta_1d \cos \theta_2d\Phi dt}\propto \sum^6_{i=1}K_i(t)f_i(\theta_1,\theta_2,\Phi) \, ,
\label{PDFpp}
\end{equation}
where the helicity angles $\Omega = (\theta_1, \theta_2, \Phi)$ are defined in
Fig.~\ref{decay_angles}. The angular functions $f_i(\Omega)$ are \cite{PhysRevD.61.074031}
\begin{eqnarray}
f_1(\theta_1,\theta_2,\Phi) &= &4\cos^2\theta_1\cos^2\theta_2 \, ,  \nonumber
\\
f_2(\theta_1,\theta_2,\Phi) &=& \sin^2\theta_1\sin^2\theta_2(1+\cos2\Phi) \, ,
\nonumber \\
f_3(\theta_1,\theta_2,\Phi) &=& \sin^2\theta_1\sin^2\theta_2(1-\cos2\Phi) \, ,
\nonumber \\
f_4(\theta_1,\theta_2,\Phi) &=& -2\sin^2\theta_1\sin^2\theta_2\sin2\Phi \, ,
\nonumber \\
f_5(\theta_1,\theta_2,\Phi) &=&
\sqrt{2}\sin2\theta_1\sin2\theta_2\cos\Phi \, , \nonumber  \\
f_6(\theta_1,\theta_2,\Phi) &=&
-\sqrt{2}\sin2\theta_1\sin2\theta_2\sin\Phi \, .
\label{feq} 
\end{eqnarray}
The time-dependent functions $K_i(t)$ are given by \cite{PhysRevD.62.014017}
\begin{eqnarray}
K_1(t)&=&\frac{1}{2}A_0^2[(1+\cos\phisphiphi)e^{-\Gamma_{\rm{L}}t}+(1-\cos\phisphiphi)e^{-\Gamma_{\rm{H}}t}
\pm2e^{-\Gamma_st}\sin(\Delta m_st)\sin\phisphiphi] \, , \nonumber \\
K_2(t)&=&\frac{1}{2}A_\parallel
^2[(1+\cos\phisphiphi)e^{-\Gamma_{\rm{L}}t}+(1-\cos\phisphiphi)e^{-\Gamma_{\rm{H}}t}\pm2e^{-\Gamma_st}\sin(\Delta
m_st)\sin\phisphiphi] \, , \nonumber \\
K_3(t)&=&\frac{1}{2}A_\perp ^2[(1-\cos\phisphiphi)e^{-\Gamma_{\rm{L}}t}+(1+
\cos\phisphiphi)e^{-\Gamma_{\rm{H}}t}\mp2e^{-\Gamma_st}\sin(\Delta
m_st)\sin\phisphiphi] \, , \nonumber \\
K_4(t)&=&|A_\parallel ||A_\perp|[\pm e^{-\Gamma_st}\{\sin\delta_1
\cos(\Delta m_st)-\cos\delta_1\sin(\Delta m_st)\cos\phisphiphi \} 
\nonumber \\
& &
- \frac{1}{2}(e^{-\Gamma_{\rm{H}}t}-e^{-\Gamma_{\rm{L}}t})\cos\delta_1\sin\phisphiphi]
\nonumber \, , \\
K_5(t)&=&\frac{1}{2}|A_0
||A_\parallel|\cos(\delta_2-\delta_1)\nonumber \\ & & [(1+\cos\phisphiphi)e^{-\Gamma_{\rm{L}}t}+(1-\cos\phisphiphi)e^{-\Gamma_{\rm{H}}t}
 \pm2e^{-\Gamma_st}\sin(\Delta m_st)\sin \phisphiphi] \, ,
\nonumber \\
K_6(t)&=&|A_0||A_\perp|[\pm e^{-\Gamma_st}\{\sin\delta_2 \cos(\Delta
m_st)-\cos\delta_2\sin(\Delta m_st)\cos\phisphiphi \} \nonumber \\ & & -\frac{1}{2}(e^{-\Gamma_{\rm{H}}t}-e^{-\Gamma_{\rm{L}}t})\cos\delta_2\sin\phisphiphi] \, , \label{keqt}
\end{eqnarray}
where the upper of the $\pm$ or $\mp$  signs refers to the \Bs meson and the lower refers to a \Bsb meson.
Here, $\GL$ and $\GH$ are the decay widths of the light and
heavy $\Bs$ mass eigenstates,\footnote{Units are adopted such that $\hbar=1$.} $\Delta m_s$ is the $\Bs$ oscillation frequency, $\delta_1=\arg(A_\perp/A_\parallel)$ and $\delta_2=\arg(A_\perp/A_0)$ are \CP-conserving strong phases and $\phisphiphi$ is the weak \CP-violating
phase. It is assumed that the weak phases of the three polarization amplitudes are equal. 
The quantities $\GH$ and $\GL$ correspond to the
observables $\DGs = \GL - \GH $ and $ \Gs = (\GL + \GH)  /2$.
In the Standard Model, the value of $\phisphiphi$ for this mode is expected
to be very close to zero due to a cancellation between the phases arising from mixing
and decay~\cite{Raidal:2002ph}.\footnote{The convention used in this Letter
is that the symbol $\phisphiphi$ refers solely to the weak phase difference measured in the $\Bs \to \phi \phi$ decay.} A calculation based on QCD factorization
provides an upper limit of 0.02 rad for $\phisphiphi$~\cite{Bartsch:2008ps,beneke}.
This is different to the
situation in the $B^0_s \rightarrow J/\psi \phi$ decay, where the
Standard Model predicts $\phi_s(J/\psi \phi) =-2\arg\left(-V_{ts}V_{tb}^*/V_{cs}V_{cb}^*\right)=-0.036\pm0.002$~rad~\cite{Charles:2011va}.  
The magnitude of both weak phase differences can be enhanced in the presence of new physics in $\Bs$ mixing,
where recent results from \lhcb have placed stringent constraints~\cite{phispaper}.
For the $\Bs \rightarrow \phi \phi$ decay, new particles could also contribute  in $b\to s$ penguin loops.

To measure the polarization amplitudes, a time-integrated untagged analysis is performed, assuming that an equal number of $\Bs$
and $\bar{B}_s^0$ mesons are produced and that the \CP-violating phase is
zero as predicted in the Standard Model.\footnote{In the case of non-zero $\phisphiphi$ deviations from these formulas are suppressed by a factor of $\DGs/\Gs$ and hence
only small variations would be observed on the fitted parameters.} In this case, the functions $K_i(t)$ integrate to
\begin{eqnarray}
K_1 &=& |A_0|^2/\GL \nonumber \, , \\
K_2 &=& |A_\parallel|^2/ \GL \, , \nonumber \\
K_3 &=& |A_\perp|^2 /\GH \, , \nonumber \\
K_4 &=& 0 \, , \nonumber \\
K_5 &=& |A_0||A_{\parallel}|\cos(\delta_\parallel) /\GL \, , \nonumber \\
K_6 &=& 0 \, ,
\label{keq}
\end{eqnarray}
where the strong phase difference is defined by $\delta_\parallel \equiv \delta_2-\delta_1 = \arg(A_\parallel/A_0)$ and the time integration assumes uniform time acceptance.

In addition, a search for physics beyond the Standard Model is performed by studying the triple product asymmetries \cite{gronau, Bensalem:2000hq, Datta:2003mj} in the $\Bs \to \phi \phi$ decay.  Non-zero values
of these quantities can be either due to $T$-violation or final-state
interactions. Assuming \CPT conservation, the former case  implies that \CP is
violated.
Experimentally, the extraction of the triple product asymmetries is straightforward and provides a measure of \CP violation that does not require flavour tagging or a time-dependent analysis.

There are two observable triple products denoted $
 U=\sin (2 \Phi)/2$ and $V=\pm \sin(\Phi)$, where the positive sign is
 taken if the $T$-even quantity $\cos \theta_1 \cos \theta_2 \geq 0$ and the negative sign
 otherwise. These variables 
correspond to the $T$-odd triple products
\begin{eqnarray}
\sin \Phi &=& (\hat{n}_{1} \times \hat{n}_{2}) \cdot \hat{p}_{1} \, ,  \nonumber \\
\sin (2\Phi)/2 &=& (\hat{n}_{1} \cdot \hat{n}_{2})(\hat{n}_{1} \times \hat{n}_{2}) \cdot \hat{p}_{1} \, ,
\end{eqnarray}
where $\hat{n}_{i}$ ($i = 1,2$) is a unit vector perpendicular to the $\phi_i$ decay 
plane and $\hat{p}_{1}$ is a unit vector in the direction of the $\phi_1$ momentum in the $\Bs$ rest frame. 
The triple products, $U$ and $V$, are proportional to the $f_4$ and $f_6$ angular functions which,  for $\phisphiphi = 0$,  vanish in the
untagged decay rate for any value of $t$. The $f_4$ and $f_6$ angular functions would not vanish in the presence of new physics processes that cause the polarization amplitudes to have different weak phases \cite{gronau}. Therefore, a measurement of significant asymmetries would be an unambiguous signal for the effects of new physics~\cite{gronau, Datta:2003mj}.

The asymmetry, $A_U$, is defined as
\begin{equation}
A_U = \frac{N_+ - N_-}{N_+ + N_-} \, ,
\label{eq:au}
\end{equation}
where $N_+$ ($N_-$) is the number of events with $U > 0$ ($U < 0$).
Similarly $A_V$ is defined as
\begin{align}
A_V = \frac{M_+ - M_-}{M_+ + M_-} \, ,
\label{eq:av}
\end{align}
where $M+$ ($M_-$) is the number of events with $V > 0$ ($V < 0$).
The triple product asymmetries, $A_U$ and $A_V$ are proportional to
the interference terms $\mathcal{I}m(A_\perp A_\parallel^*)$ and $\mathcal{I}m(A_\perp A_0^*)$
in the decay rate.

The $\Bs \rightarrow \phi \phi$ decay mode was first observed by the CDF
Collaboration~\cite{Acosta:2005eu}. More recently, CDF has reported
measurements of the polarization amplitudes and triple product
asymmetries in this mode based on a sample of 295 events
\cite{cdfpaper}. 
In this Letter, measurements of the polarization amplitudes, $|A_0|^2$ and $|A_\perp|^2$,
the strong phase difference, $\delta_\parallel$,  and the triple product asymmetries,
$A_U$ and $A_V$, are presented.
The dataset consists of $801 \pm 29$ candidates collected in $1.0~\invfb$ of $pp$ collisions at the LHC.
The Monte Carlo (MC) simulation samples used are based on the
{\sc{Pythia}}~6.4 generator~\cite{pythia} configured with the
parameters detailed in Ref.~\cite{belyaev}. The
{\sc{EvtGen}}~\cite{evtgen} and  {\sc{Geant4}}~\cite{geant} packages
are used to generate hadron decays and simulate interactions in the detector, respectively. 

\section{Detector description}
\label{sec:Detector}
The \lhcb detector~\cite{LHCbDetector} is a single-arm forward
spectrometer covering the pseudorapidity range $2<\eta <5$, designed
for the study of particles containing \bquark or \cquark quarks. The
detector includes a high precision tracking system consisting of a
silicon-strip vertex detector surrounding the $pp$ interaction region,
a large-area silicon-strip detector located upstream of a dipole
magnet with a bending power of about $4{\rm\,Tm}$, and three stations
of silicon-strip detectors and straw drift-tubes placed
downstream. The combined tracking system has a momentum resolution
$\Delta p/p$ that varies from 0.4\% at 5\gevc to 0.6\% at 100\gevc,
and an impact parameter resolution of 20\mum for tracks with high
transverse momentum. Charged hadrons are identified using two
ring-imaging Cherenkov detectors. Photon, electron and hadron
candidates are identified by a calorimeter system consisting of
scintillating-pad and pre-shower detectors, an electromagnetic
calorimeter and a hadronic calorimeter. Muons are identified by a muon
system composed of alternating layers of iron and detector stations. The trigger consists of a hardware stage, based
on information from the calorimeter and muon systems, followed by a
software stage which applies a full event reconstruction.

The software trigger used in this analysis requires a two-, three- or four-track
secondary vertex with a high sum of the transverse momentum, \pt, of
the tracks, significant displacement from the primary interaction,
and at least one track with $\pt > 1.7\gevc$; impact parameter
\chisq with respect to the primary interaction greater than 16; and
a track fit $\chisq/\rm{ndf} < 2$ where ndf is the number of degrees of freedom in the track fit. A multivariate algorithm is used
for the identification of the secondary
vertices~\cite{LHCb-PUB-2011-016}.
The $\Bs \rightarrow \phi \phi$ candidates are
selected with high efficiency either by identifying
events containing a $\phi$ meson or using topological
information to select hadronic \bquark decays. Events passing the software trigger are
stored for subsequent offline processing.

\section{Event selection} 
\label{sec:selection}
The $\Bs \rightarrow \phi \phi$ channel is reconstructed using events
where both $\phi$ mesons decay into a $K^+K^-$ pair.
The $\Bs \rightarrow \phi \phi$ selection criteria were optimized using a data-driven
approach based on the $\phantom{}_s\mathcal{P}lot$ technique employing the four-kaon mass as the unfolding variable \cite{Pivk:2004ty} to
separate signal ($S$) and background ($B$) with the aim of maximizing
$S/\sqrt{S+B}$. The resulting cuts are summarized in
Table~\ref{selecphiphi}. Good quality track reconstruction is ensured by a cut
on the transverse  momentum (\pt) of the daughter particles and a
cut on the $\chi^2/{\rm ndf}$ of the track fit. 

Combinatorial background is reduced by cuts on the minimum impact
parameter significance of the tracks with respect to all reconstructed
$pp$ interaction vertices and also by imposing a requirement
on the vertex separation $\chi^2$ of the $\Bs$ candidate. Well-identified $\phi$ meson
candidates are selected by requiring that the two particles involved are
identified as kaons by the ring-imaging Cherenkov detectors using a cut on the difference in the
global likelihood between the kaon and pion hypotheses ($\DLL{\mathit{K}}{\mathit{\pi}} > 0$) and
  by requiring that the reconstructed mass of each $K^+K^-$ pair is within $12~\MeVcc$
of the nominal mass of the $\phi$ meson \cite{Nakamura:2010zzi}. Further signal purity is
achieved by cuts on the transverse momentum of the $\phi$ candidates.
\begin{table}[t]
\caption{Selection criteria for the $\Bs \rightarrow \phi
  \phi$ decay. The abbreviation IP stands for impact parameter and $\pt^{\phi1}$ and $\pt^{\phi2}$
  refer to the transverse momentum of the two $\phi$ candidates.}
\begin{center}
\begin{tabular}{ll}
\hline
Variable & Value \\ \hline
Track $\chi^2/{\rm ndf}$ & $<~5$ \\
Track $\pt$    & $>~500~\MeVc$  \\
Track IP $\chi^2$ & $>~21$\\
$\DLL{\mathit{K}}{\mathit{\pi}}$ & $>~0$\\
$|M_{\phi}-M_{\phi}^{\rm{PDG}}|$ &  $<~12~\MeVcc$\\
$\pt^{\phi1}, \, \pt^{\phi2}$ & $>~900~\MeVc$  \\
$\pt^{\phi1}\cdot\pt^{\phi2}$ & $>~2$~GeV$^2/c^2$  \\
$\phi$ vertex $\chi^2/{\rm ndf}$  & $<~24$\\
$\Bs$ vertex $\chi^2/{\rm ndf}$  & $<~7.5$ \\
$\Bs$ vertex separation $\chi^2$  & $>~270$ \\
$\Bs$ IP $\chi^2$ & $<~15$ \\
\hline
\end{tabular}
\label{selecphiphi}
\end{center}
\end{table}

\begin{figure}[t]
\begin{center}
\includegraphics[height=8cm]{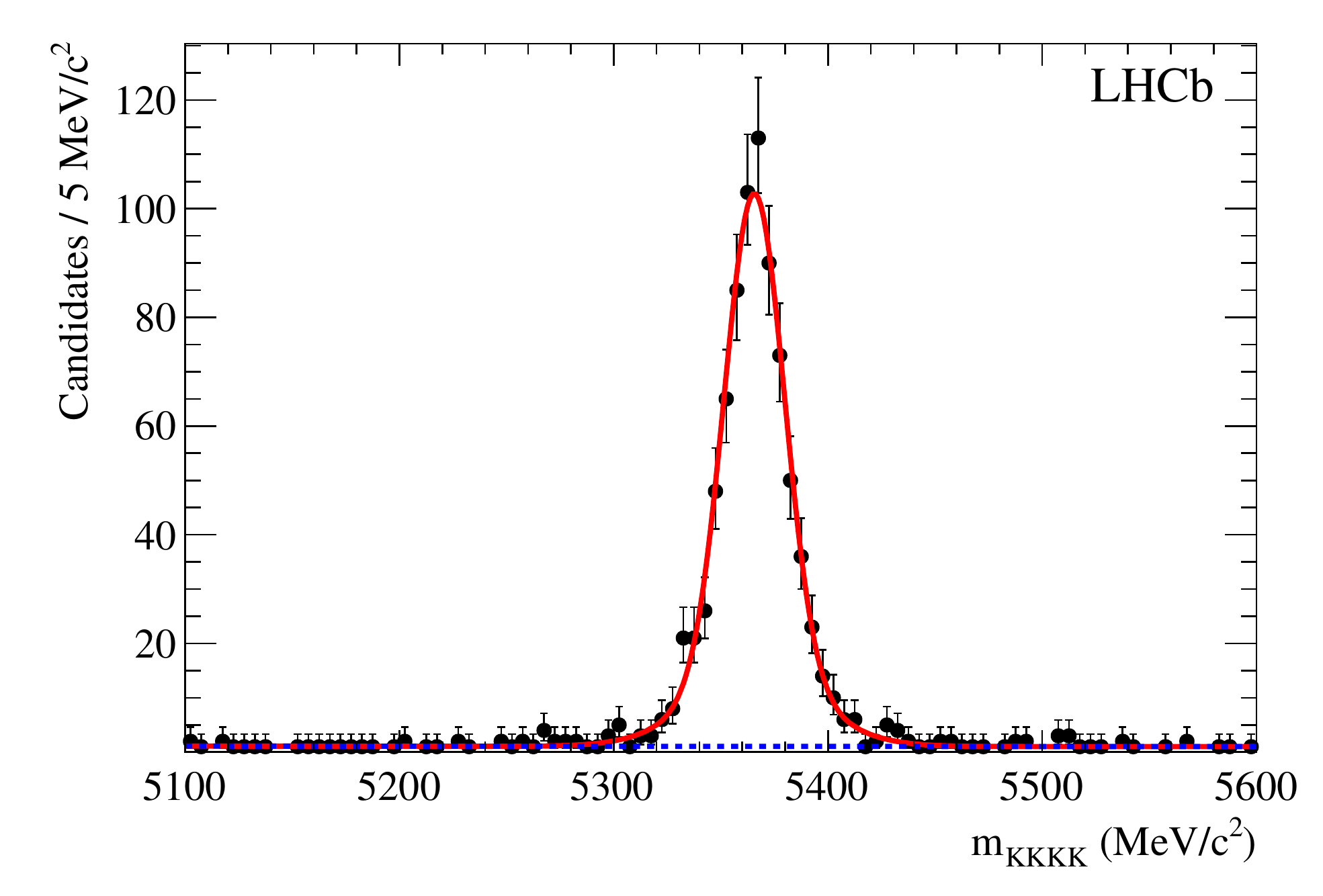}
\end{center}
\caption{Invariant $K^+K^-K^+K^-$ mass distribution for selected $\Bs \rightarrow \phi \phi$
  candidates. A fit of a double
  Gaussian signal component together with an exponential background
  (dotted line)  is superimposed.}
\label{ppmass}
\end{figure}

Figure~\ref{ppmass} shows the four-kaon invariant mass distribution for selected
events. To determine the signal yield an unbinned maximum likelihood fit is 
performed. The $\Bs \rightarrow\phi \phi$ signal component is modelled by two
Gaussian functions with a common mean. The resolution of the first Gaussian is measured from data to be 
$13.9 \pm 0.6$~\MeVcc.
The relative fraction and resolution
of the second Gaussian are fixed to 0.785 and 29.5~\MeVcc respectively, where values have been obtained from simulation. Combinatorial background is modelled 
using an exponential function. Background from $\Bd \rightarrow \phi K^{*0}$ and $\Bs
\rightarrow K^{*0} \Kstarzb$ decays is found to be negligible both in simulation and data
driven studies. Fitting the probability density function (PDF) described above to the data, a signal yield of
$801 \pm 29$ events is found.

In addition to the dominant P-wave $\phi \rightarrow K^+K^-$ component
described in Section~\ref{sec:Introduction}, other contributions,
either from  $f_0 \rightarrow K^+K^-$ or non-resonant $K^+
K^-$, are possible. The size of these contributions,
neglecting interference effects, is studied by relaxing the $\phi$ mass cut
to be within $25$~\MeVcc of the nominal value\footnote{This is a larger window than the 
$\pm12$~\MeVcc window used in the polarization amplitude and strong phase difference measurements.} 
and using the $\phantom{}_s\mathcal{P}lot$ technique in conjunction with the $\phi$ mass to subtract the combinatorial background.

\begin{figure}[h!]
\begin{center}
\resizebox{3.8in}{!}{\includegraphics{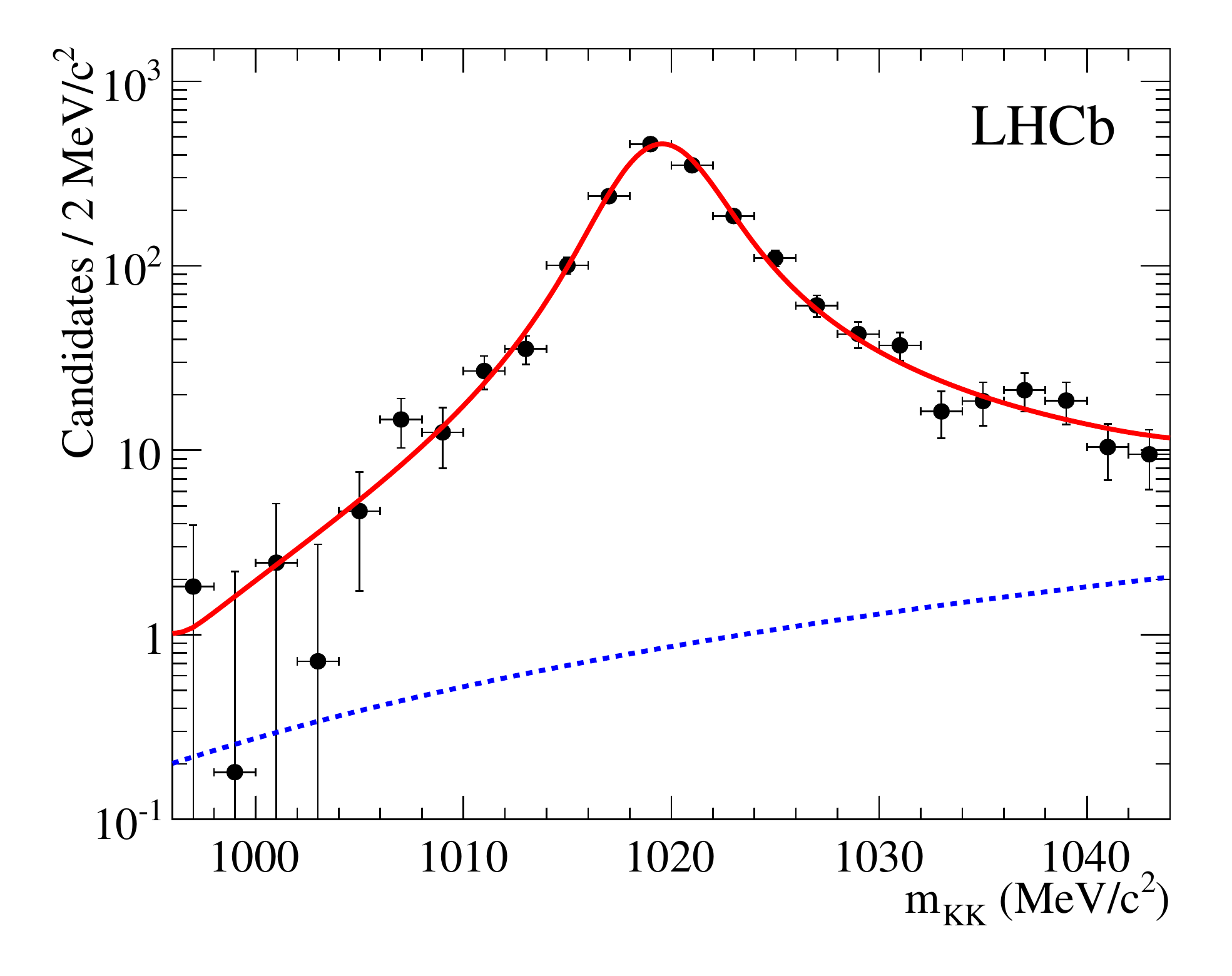}}
\caption{ \label{fig:phiphisplot.pdf} Invariant mass distribution of $K^+K^-$ pairs for the $B^0_s \rightarrow
  \phi \phi$ data without a $\phi$ mass cut. The background
  has been removed using the $_{s}\mathcal{P}lot$ technique in conjunction with the $K^+K^-$ invariant mass. There are two entries per $B^0_s$ candidate. The solid line
  shows the result of the fit model described in the text. The fitted S-wave
  component is shown by the dotted line.}
\end{center}
\end{figure}

The resulting $\phi$ mass distribution is shown in
Fig.~\ref{fig:phiphisplot.pdf}. A fit of a relativistic P-wave Breit-Wigner function together with a two body phase space component to model the S-wave
contribution is superimposed. In a $ \pm 25~\MeVcc$ mass window, the size of the S-wave component is found to be $(1.3 \pm
1.2)\%$. Since the S-wave yield is
consistent with zero, it will be neglected in 
the following section. A systematic uncertainty arising from this
assumption will be assigned.

\section{Results}
\label{sec:triplestudies}
\begin{figure}[h]
\centering
\subfigure{
\includegraphics[height=4.5cm]{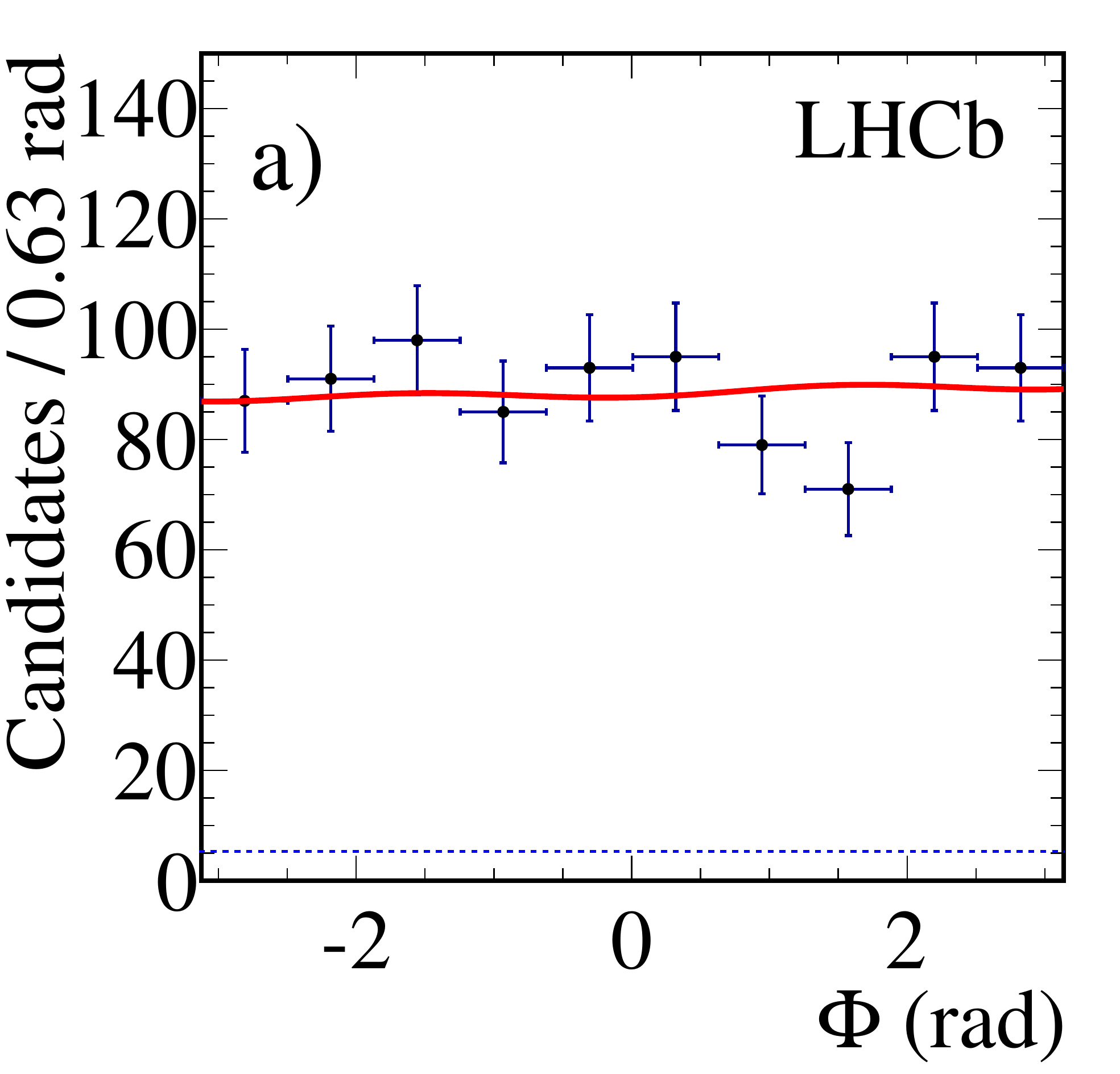}
\label{fig:pDTphi}
}
\subfigure{
\includegraphics[height=4.5cm]{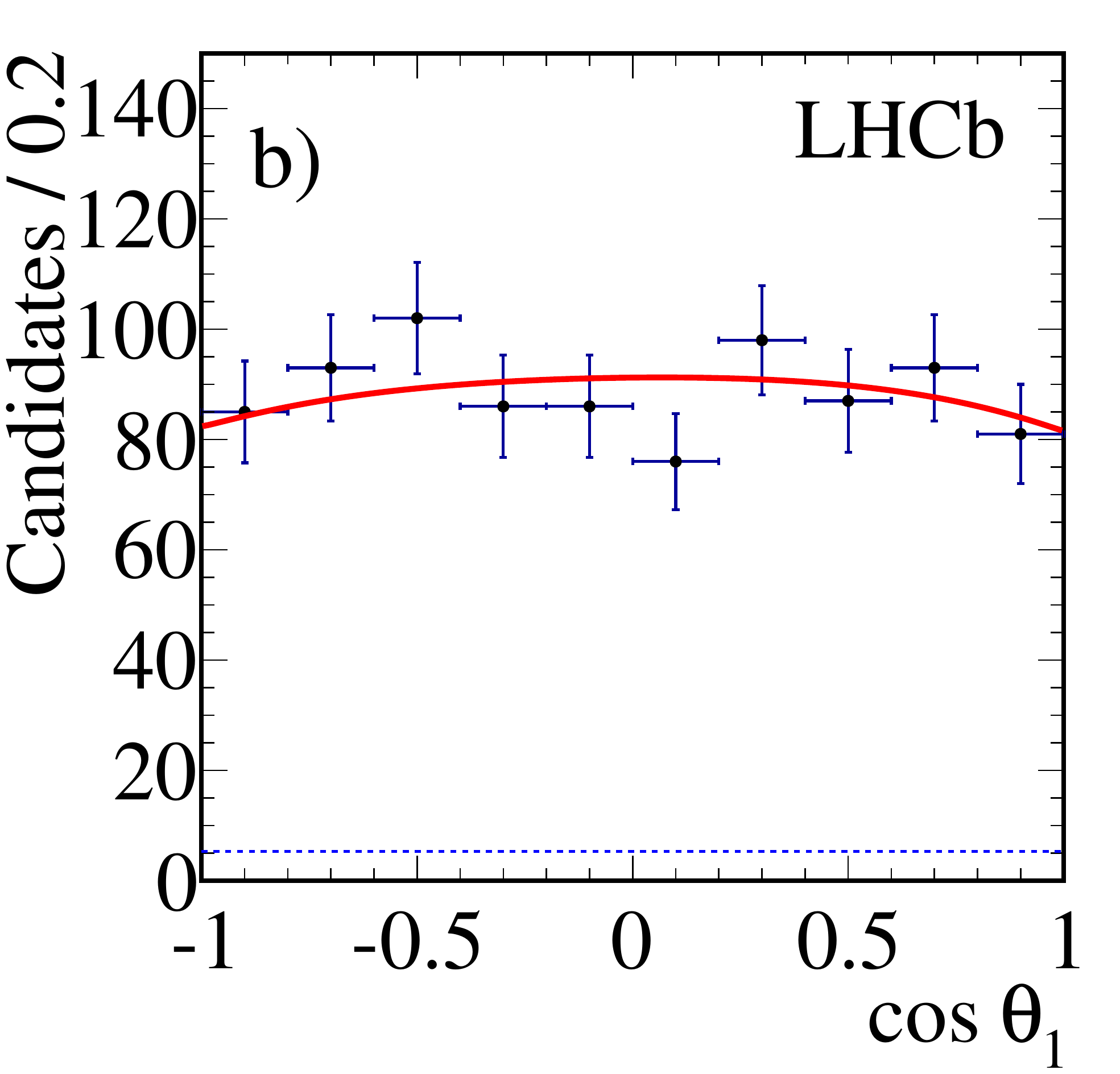}
\label{fig:pDTCT1}
}
\subfigure{
\includegraphics[height=4.5cm]{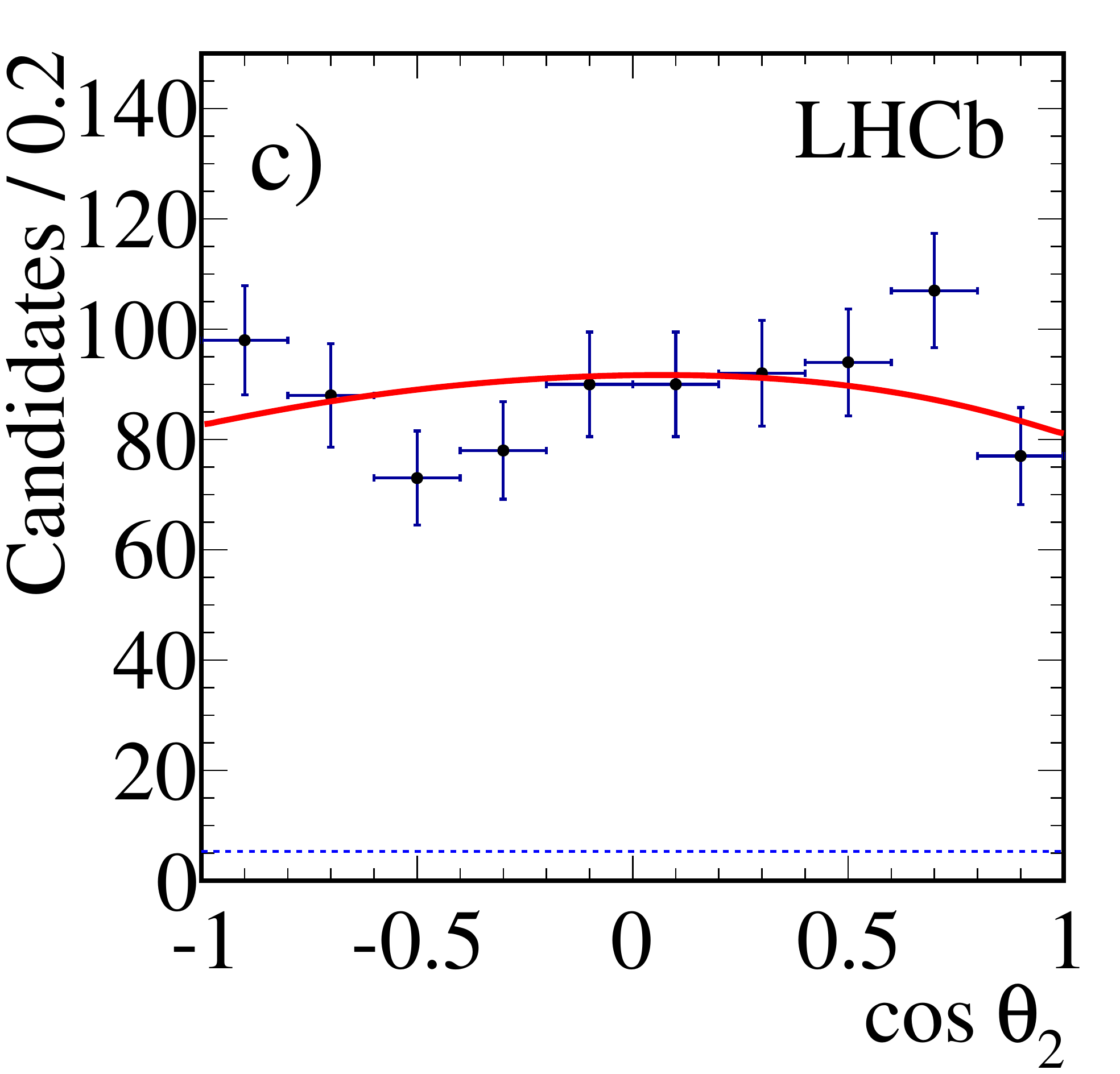}
\label{fig:pDTCT2}
}
\caption{ Angular distributions for (a) $\Phi$, (b) $\cos\theta_1$ and (c) $\cos\theta_2$ of $B_s^0 \rightarrow \phi \phi$
  events with the fit projections for signal and background superimposed for the
  total fitted PDF (solid line) and background
  component (dotted line).}
\label{TIDataC_proj}
\end{figure}

The polarization amplitudes ($|A_0|^2$, $|A_\perp|^2$, $|A_\parallel|^2$), are determined by performing an unbinned
maximum likelihood fit to the reconstructed mass and helicity
angle distributions. For each event, the $\phi$ meson used to define $\theta_1$ is chosen at random. Both the signal and background PDFs are the products of a mass
component described in Section~\ref{sec:selection} together with an angular
component. The angular component of the signal is given by Eq.~\ref{feq} multiplied
by the angular acceptance of the detector. The acceptance is determined
using the simulation and is calculated separately according to trigger type, i.e.
whether the event was triggered by the signal candidate or other
particles in the event. 
In total the fit 
for the polarization amplitudes has eight free parameters: the 
signal angular parameters $|A_0|^2$, $|A_\perp|^2$ and 
$\cos(\delta_{\parallel})$ defined in Section~\ref{sec:Introduction}, 
the fractions of signal for each trigger type, the 
resolution of the core Gaussian, the $B^{0}_s$ mass and the slope of the mass
background. The sum of
squared amplitudes is constrained such that $|A_0|^2+|A_\perp|^2+|A_\parallel|^2=1$. 
The angular distributions for the
background have been studied using the mass sidebands in the data, where mass sidebands are defined to be between $60$ and $300$~\MeVcc either side of the nominal \Bs mass \cite{Nakamura:2010zzi}. With the
current sample size these distributions are consistent with being flat
in ($\cos\theta_1, \cos\theta_2, \Phi$). Therefore, a uniform angular PDF is assumed and more
complicated shapes are considered as part of the systematic
studies. The values of $\Gamma_s = 0.657 \pm 0.009 \pm 0.008 \ps^{-1}$
and $\Delta \Gamma_s = 0.123 \pm 0.029 \pm 0.011 \ps^{-1}$ together with their correlation coefficient of $-0.3$ quoted in \cite{phispaper} are used as a Gaussian constraint.
The validity of the fit model has been extensively tested using simulated data
samples. The results are given in Table~\ref{resComb} and the
angular projections are shown in Fig.~\ref{TIDataC_proj}. 
\begin{table}[h]
\caption{Measured polarization amplitudes and strong phase difference. The
  uncertainties are statistical only. The sum of the
  squared amplitudes is constrained to unity. The correlation coefficient
  between ${|A_0|}^2$  and  ${|A_{\perp}}|^2$ is $-0.47$.}
\begin{center}
\begin{tabular}{lr@{$\pm$}l} 
\hline
           Parameter &  \multicolumn{2}{c}{Measurement}                    \\ \hline
           ${|A_0|}^2$ &        0.365 &      0.022                \\
       ${|A_\perp|}^2$ &         0.291 &    0.024                \\
       ${|A_\parallel|}^2 = 1 - (|A_0|^2 + |A_\perp|^2)$ &         0.344 &    0.024                \\
  $\cos(\delta_\parallel)$ &       $-$0.844 &     0.068                \\
  \hline
\end{tabular}
\label{resComb}
\end{center}
\end{table}

Several sources of systematic uncertainty on the determination of the polarization
amplitudes are considered and summarized in Table~\ref{systAmp}. With the present size of the dataset, the S-wave component is consistent with zero. From the studies
described in Section~\ref{sec:selection} and fits to the data
including the S-wave terms in the PDF \cite{Xie:2009fs}, we consider a maximum S-wave component of $2\%$.
Simulation studies have been performed to investigate the effect of neglecting an S-wave
component of this size. As discussed in Section~\ref{sec:Introduction}, the integration that leads to Eq.~\ref{keq} assumes uniform time acceptance. This is not the case due to lifetime biasing cuts in the trigger and offline selections. The functional form of the decay time acceptance is obtained through
the use of Monte Carlo events. The difference between using this functional form in simulation studies and using uniform time acceptance is taken as a systematic uncertainty. The uncertainty on the angular acceptance for the signal is
propagated to the observables also using Monte Carlo studies.
The analysis was repeated with an alternative background angular distribution, taken from a coarsely binned histogram in $(\cos \theta_1,\cos\theta_2, \Phi)$ of the mass sidebands, and the difference taken as a systematic uncertainty. An additional uncertainty arises from angular acceptance dependencies on trigger
type. This dependency is corrected for using Monte Carlo events, with half of the effect
on fitted parameters assigned as systematic uncertainties.
The total systematic uncertainty is obtained from the sum in quadrature of the individual uncertainties.

\begin{table}
\caption{Systematic uncertainties on the measured polarization amplitudes and the strong phase difference.}
\begin{center}
\begin{tabular}{l|c|c|c|c} 
\hline
Source &  $|A_0|^2$ & $|A_\perp|^2$ & $|A_\parallel|^2$ & $\cos\delta_\parallel$  \\
\hline 
S-wave component & 0.007 & 0.005 & 0.012 & 0.001 \\ 
Decay time acceptance  & 0.006 & 0.006 & 0.002 & 0.007 \\
Angular acceptance & 0.007 & 0.006  & 0.006  & 0.028 \\
Trigger category & 0.003 & 0.002 & 0.001 & 0.004 \\
Background model & 0.001 & -- & 0.001 & 0.003 \\ \hline
Total & 0.012 & 0.010 & 0.014 &  0.029 \\
\hline
\end{tabular}
\label{systAmp}
\end{center}
\end{table}

\begin{figure}[h]
\includegraphics[width=0.5\textwidth]{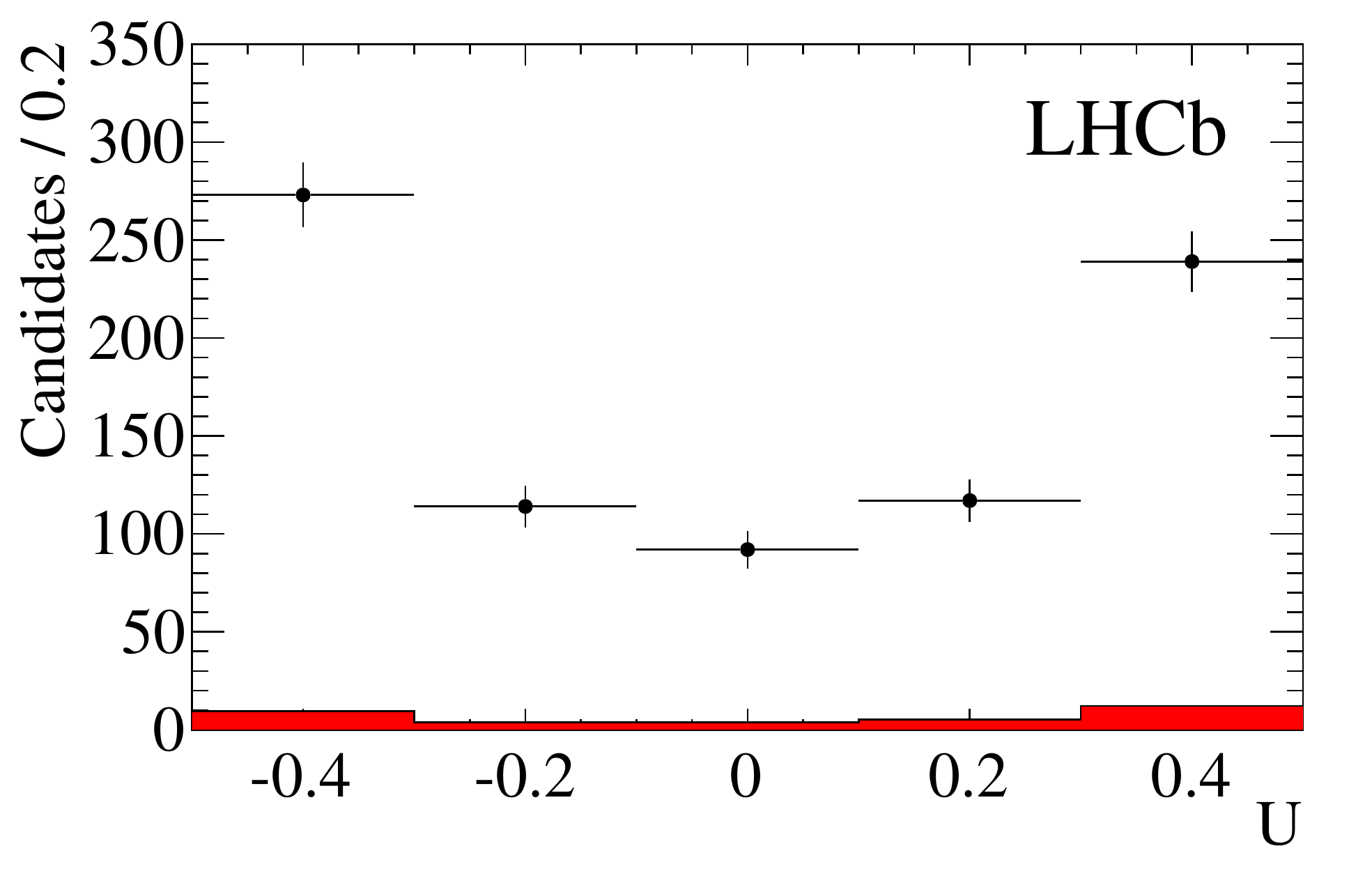}
\includegraphics[width=0.5\textwidth]{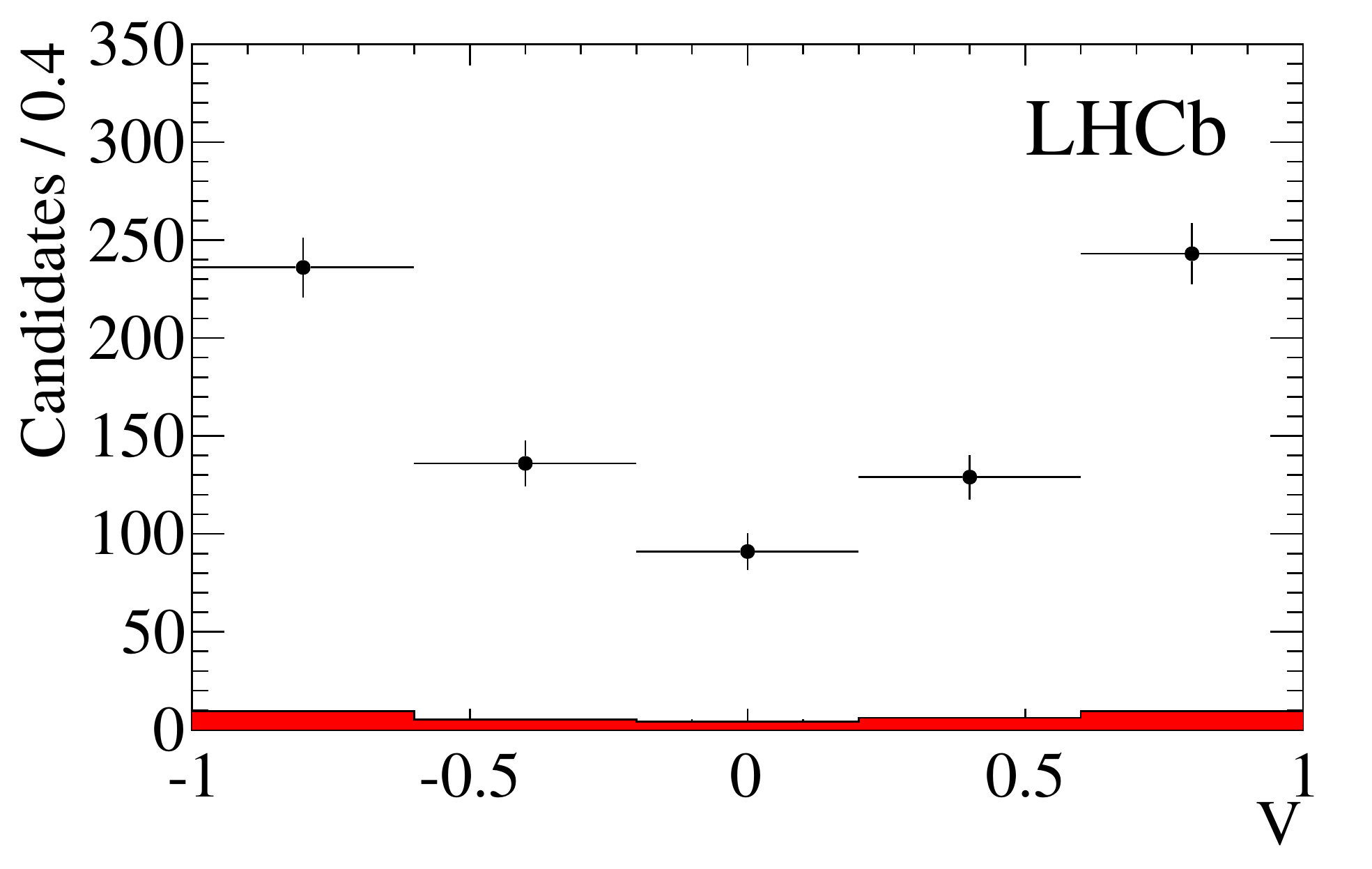}
\caption{Distributions of the $U$ (left) and $V$ (right) observables for the $B_s^0
  \rightarrow \phi \phi$ data in the mass range $5286.6 < M(K^+K^-K^+K^-) <
  5446.6~\MeVcc$.  The distribution for the background is taken from the
  mass sidebands, normalized to the same mass range and is shown by
  the solid histogram.}
\label{UVobs}
\end{figure}

The distributions of the $U$ and $V$ triple product observables are shown in
Fig.~\ref{UVobs} for the mass range  $5286.6 < M(K^+K^-K^+K^-) < 5446.6~\MeVcc$. 
To determine the triple product asymmetries, the dataset is partitioned according to whether $U$ ($V$) is less than
or greater than zero. Simultaneous fits are performed to the mass
distributions for each of the two partitions corresponding to each observable individually. In these fits, the mean and
resolution of the Gaussian signal component together with the slope of the
exponential background component are common parameters. The asymmetries are left as free parameters and are fitted for directly in the simultaneous fit. The measured values are
\begin{center}
\begin{tabular}{l@{$~=~$}r@{$\,\,\pm\,$}l}
$A_U$ & $-$0.055 & 0.036 $\, ,$  \\
$A_V$ & 0.010 & 0.036 $\, .$
\end{tabular}
\end{center}

Systematic uncertainties due to the residual effect of the decay time, geometrical acceptance and the signal and background fit models have been evaluated and are summarized in
Table~\ref{systtable}. The effect of the decay time acceptance has been found using the same method as 
for the polarization amplitudes. The impact of angular acceptance on the measured values has been obtained from simplified simulation
studies. The total systematic uncertainty is
conservatively estimated by choosing the larger of the two individual systematic uncertainties on  $A_U$ and
$A_V$. The contributions are combined
in quadrature to determine the total systematic error. Various
cross-checks of the stability of the result have been performed. For
example, dividing the data according to how the event was triggered or by magnet polarity. No
significant bias is observed in these studies.

\begin{table}[htb]
\begin{center}
\caption[]{  {\label{systtable}}  Systematic
  uncertainties on the triple product asymmetries $A_U$ and $A_V$. The total uncertainty is the quadratic sum of the larger of the two components.}
\begin{tabular}{l|c|c|c}
\hline
Source & $A_U$ & $A_V$ & Final uncertainty \\ \hline
Angular acceptance  & 0.009 & 0.006 & 0.009 \\
Decay time acceptance  & 0.006 & 0.014 & 0.014 \\
Fit model & 0.004 & 0.005 & 0.005 \\ \hline
Total  &   \multicolumn{2}{c}{}   & $0.018$   \\
\hline
\end{tabular}
\end{center}
\end{table}

\section{Summary}  
\label{sec:summary}
The polarization amplitudes and strong phase difference
in the $\Bs \rightarrow \phi \phi$ decay mode are measured to be
\begin{center}
\begin{tabular}{l@{$~=~$}r@{$\,\pm\,$}l@{$\,$(stat) 
      $\pm\,$}l@{$\,$(syst)}l@{\,}l}
$|A_0|^2$ & 0.365 & 0.022 & 0.012  & $\, ,$ \\
$|A_\perp|^2$  & 0.291 & 0.024 & 0.010 & $\, ,$  \\
$|A_\parallel|^2 $ & 0.344 & 0.024 & 0.014& $\, ,$ \\
$\cos(\delta_\parallel)$ &  $-$0.844 &  0.068 & 0.029 & $\, ,$
\end{tabular}
\end{center}
where the sum of the squared amplitudes is constrained to be unity. These values agree well with the CDF measurements
\cite{cdfpaper}. Measurements in other $B \rightarrow VV$ penguin transitions at the
\textit{B} factories generally give higher values of $f_L$~\cite{PhysRevLett.94.221804,PhysRevLett.98.051801,PhysRevD.78.092008, delAmoSanchez:2010mz,Abe:2004mq, Aubert:2006fs}. 
It is interesting to note that the value of $f_L$ found in the $\Bs \rightarrow
\phi \phi$ channel is almost equal to that in
the $B^0_s \rightarrow K^{*0} \Kstarzb$ decay \cite{kstarlhcb}. The 
results are in agreement with QCD factorization predictions
\cite{beneke, PhysRevD.80.114026}, but disfavour the pQCD estimate given in \cite{PhysRevD.76.074018}. 

The triple product asymmetries in this mode are measured to be
\begin{center}
\begin{tabular}{l@{$~=~$}r@{$\,\pm\,$}l@{$\,$(stat)
      $\pm\,$}l@{$\,$(syst)}l@{\,}l}
$A_U$ & $-$0.055 & 0.036 & 0.018 & $\, ,$ \\
$A_V$ & 0.010 & 0.036 & 0.018 & $\, .$
\end{tabular}
\end{center}
Both values are in good agreement with those reported by the CDF
Collaboration \cite{cdfpaper} and  consistent
with the hypothesis of \CP conservation.

\section*{Acknowledgements}

\noindent We express our gratitude to our colleagues in the CERN accelerator
departments for the excellent performance of the LHC. We thank the
technical and administrative staff at CERN and at the LHCb institutes,
and acknowledge support from the National Agencies: CAPES, CNPq,
FAPERJ and FINEP (Brazil); CERN; NSFC (China); CNRS/IN2P3 (France);
BMBF, DFG, HGF and MPG (Germany); SFI (Ireland); INFN (Italy); FOM and
NWO (The Netherlands); SCSR (Poland); ANCS (Romania); MinES of Russia and
Rosatom (Russia); MICINN, XuntaGal and GENCAT (Spain); SNSF and SER
(Switzerland); NAS Ukraine (Ukraine); STFC (United Kingdom); NSF
(USA). We also acknowledge the support received from the ERC under FP7
and the Region Auvergne.

\ifx\mcitethebibliography\mciteundefinedmacro
\PackageError{LHCb.bst}{mciteplus.sty has not been loaded}
{This bibstyle requires the use of the mciteplus package.}\fi
\providecommand{\href}[2]{#2}

\end{document}